\documentclass[sn-mathphys-ay]{sn-jnl}
\usepackage{graphicx}
\usepackage{multirow}
\usepackage{amsmath}
\usepackage{amssymb}
\usepackage{amsfonts}
\usepackage{mathrsfs}
\usepackage{rotating}
\usepackage{appendix}
\usepackage{textcomp}
\usepackage{manyfoot}
\usepackage{booktabs}
\usepackage{etoolbox}
\usepackage{algorithm}
\usepackage{xcolor}
\usepackage{algorithmicx}
\usepackage{algpseudocode}
\usepackage{listings}
\usepackage{url}
\usepackage{natbib}
\usepackage{tikz}
\usetikzlibrary{positioning}

\def\spacingset#1{\renewcommand{\baselinestretch}%
{#1}\small\normalsize} \spacingset{1}

\begin{document}

\title[Article Title]{Boosting with copula-based components}

\author*[1]{\fnm{Simon Boge} \sur{Brant}}\email{simonbb@math.uio.no}
\author[1]{\fnm{Ingrid Hobæk Haff} \sur{Author}}

\affil*[1]{\orgdiv{Department of Mathematics}, \orgname{University of Oslo}, \orgaddress{\city{Oslo}, \country{Norway}}}

\abstract{
The authors propose new additive models for binary outcomes, where the components are copula-based regression models \citep{noh2013copula}, and designed such that the model may capture potentially complex interaction effects. The models do not require discretisation of continuous covariates, and are therefore suitable for problems with many such covariates. A fitting algorithm, and efficient procedures for model selection and evaluation of the components are described. Software is provided in the R-package copulaboost. Simulations and illustrations on data sets indicate that the method's predictive performance is either better than or comparable to the other methods.
}

\maketitle

\section{Introduction}\label{sec:intro}
When the main purpose of statistical modelling is to produce
predictions for new observations, rather than, for instance, to make
inferences about a population as a whole or to summarise a given data set,
there are many popular modelling techniques, such as neural networks
\citep{lecun1998gradient}, random forests \citep{breiman2001random}, or
gradient boosting \citep{friedman2001greedy}. Which modelling technique that is
the most suitable for prediction depends on the amount of data available to fit 
the model, whether the variables are discrete or continuous, how many 
variables there are, and how complex the true relationship between the 
covariates and the response are.

In some settings, interaction effects are especially important, for instance in medical applications, such as when one tries to model the relationship between genetic data, for example SNP microarray data, and a disease \citep{ruczinski2003logic}. Another example is fraud detection applications,
where one often can expect the underlying mechanisms to be complex and multifaceted, such that interaction effects are likely to be present.
In this paper, an approach to prediction that is designed to be especially well suited for settings where there are complex interactions between covariates is proposed. Although this presentation concentrates on modelling data with a binary response variable, and presents the methodology in this setting, the models can be seen to have a form that is similar to a generalised additive model (GAM) \citep{hastie1990generalized}, in that they can be written as
$$
g^{-1}\left(\mathbb{E}(Y \vert \mathbf{X} = \mathbf{x})\right) = h_0 + \sum_{m=1}^M h_m(\mathbf{x}),
$$
where $g^{-1}(t) = \log\left(\frac{t}{1-t}\right)$ in this case. Thus, an extension of the approach to settings with another type of response variable is relatively straightforward.

One modelling framework that is designed to capture complex interactions is
logic regression \citep{ruczinski2003logic, ruczinski2004exploring}. This is
inspired by applications for genetic data, where single base-pair differences
(SNPs) are used as covariates, and one suspects that there are complicated
interaction effects. In terms of form, these can be written as additive models, where each model component is a logic tree.
As each variable in such genetic data is or can be re-coded to being binary, logic regression is tailored
to data with only binary covariates. It can be extended to also include
some continuous covariates by constructing binary covariates as indicators
of whether they exceed a certain threshold. However, this is a very crude
discretisation of continuous data, which may lead to poor prediction
performance.

Other approaches well suited for situations that involve interaction effects
are gradient boosted decision trees \citep{friedman2001greedy}, and neural nets \citep{schmidhuber2015}.
Decision trees involve discretization of the covariates, although smooth effects may be approximated
by a collection of trees, this is still a weakness of decision trees. Neural nets, on the other hand, 
typically require large amounts of training data, and lack interpretability. For instance, it is not
obvious how to determine which covariates, or combinations of covariates, are influential for a certain
predicted outcome, which is of essence in many applications \citep{rudin2019}.
%
%

In order to build a modelling framework that both captures complicated
interaction effects and takes into account the whole distribution of
continuous covariates, while remaining interpretable, a pair-copula based
approach can be used to construct model components. Specifically,  one can iteratively fit joint models to the covariates and the scoring function. The components can then be derived from the joint model as the conditional expectation of the scoring function, given the covariates. The resulting model is of the same form as classic boosting models, but with model components based on pair-copula constructions instead of decision trees, which avoids the discretisation of continuous covariates. That is, in this approach the resulting model is an additive model, where each component in the model is expressed as a conditional expectation in a pair-copula model, in contrast to the work of 
\citet{vatter2018generalized}, where the parameters of a single pair-copula model depends on covariates through generalised additive models.

There are some computational challenges related to fitting and evaluation of these components, and in order to combat these, an efficient fitting procedure is developed, as well as computationally feasible procedures for model selection and component evaluation, based on a combination of model simplifications and approximations. These procedures for model selection and evaluation of the model components are not only absolutely necessary for the model to be a realistic alternative in higher dimensions, they might also be used as a basis
for designing efficient models selection algorithms for other types of copula regression models.

The rest of the paper is organised as follows. Section \ref{sec:model}
gives a presentation of our model and an introduction to pair-copula
constructions. The procedures for estimation, model selection and
evaluation of model components are presented in Section
\ref{sec:estimation}. The properties of our method are explored in
a simulation study in Section \ref{sec:sim_study}, and illustrated
on a set of breast cancer data and on the Boston housing dataset. Finally,
some concluding remarks are given in Section \ref{sec:conclusion}.

\section{Model}\label{sec:model}

The purpose of the framework described in this paper is to construct models that can capture complex interaction effects, in an
adaptive way, for both discrete and continuous covariates. The scope is restricted to a 
regression problem with a binary outcome, and the proposed model can be written
as an additive model for a logit-link transformed conditional expectation, that is fitted by adapting a gradient boosting approach (see Section \ref{sec:estimation}). More specifically, the model can be
written as 
\begin{equation}
\log\left(\frac{\text{Pr}(Y=1 \vert \mathbf{X} = \mathbf{x})}
{\text{Pr}(Y=0 \vert \mathbf{X}=\mathbf{x})}\right) = h(\mathbf{x}) =
h_0 + \sum_{m=1}^M h_m(\mathbf{x}),
\label{eqn:model}
\end{equation}
but it differs from classic boosting models as the base learners $h_m$ are based
on pair-copula constructions (PCCs) instead of decision trees. Therefore, a short
introduction to PCCs follows. Each base learner should only involve a subset
$\mathbf{x}_{S,m}$ of the covariates, and the model may in turn contain many such
components, resulting in a large $M$.

\subsection{Pair-copula constructions}\label{subsec:pccs}
A copula is a function $C: [0, 1]^d \rightarrow [0, 1],$ that has the same
properties as a cumulative distribution function of a $d$-dimensional random
variable defined on the $d$-dimensional unit hypercube
\citep{nelsen2007introduction}. Most fundamental to the theory of copulas, is
\textit{Sklar's theorem} \citep{sklar1959fonctions}, which states that for any
random vector $\mathbf{X} = (X_1, X_2, \dots, X_d)$, with joint distribution
function $H$ and marginal distribution functions $F_1, F_2, \dots, F_d$, there
exists a copula $C$ such that
$$
H(x_1, x_2, \dots, x_d) = C\left(F_1(x_1), F_2(x_2), \dots, F_d(x_d)\right).
$$
This result makes it possible to decouple the dependency structure of a random 
vector from the particular characteristics of each margin. Hence, the modelling
can be done in two separate steps: finding suitable marginal distributions for
all the $d$ variables, and specifying a copula that describes the dependence
between them.

The base learners $h_{m}$ of the model \eqref{eqn:model} are in this paper restricted to parametric copulas. The list of available bivariate copulas is long and varied. In dimensions $3$ and higher, on the other hand, the choice is much  more limited. An alternative is then a hierarchical copula model, having smaller
copulas as components. The most widely used among these is pair-copula
constructions
\citep{joe1996families, bedford2001probability, bedford2002vines, aas2009pair},
that decompose multivariate copulas into a combination of bivariate components.
All the involved pair-copulas can be selected completely freely, which makes
PCCs extremely flexible, and able to portray a wide range of complex
dependencies \citep{niko-2008}. Inference on PCCs is in general demanding, but
there are several subclasses with appealing computational properties, the
largest of which is called regular vines (R-vines) introduced by
\cite{bedford2002vines}, and described in more detail in \cite{kurowicka11}. In
this paper, however, a smaller class of PCCs called D-vines is used,
which again is a subclass of regular vines. The reason for this limitation is that the properties of this class allows for the design of an efficient greedy algorithm for selecting the subsets $S$ of covariates that the model components should be composed of, similar to the approach taken by \citet{schallhorn2017d} for quantile regression.

\begin{figure}
  \centering
\begin{tikzpicture}[
    roundnode/.style={circle, draw=green!60, fill=green!5, very thick,
      minimum size=7mm},
]
\node[roundnode]      (var1)                              {$1$};
\node[roundnode]        (var2)       [right=of var1] {$2$};
\node[roundnode]      (var3)       [right=of var2] {$3$};
\node[roundnode]        (var4)       [right=of var3] {$4$};

\draw[-] (var1.east) -- (var2.west) node[scale =0.75, pos=0.5, above, align=left] {$1, 2\,$};
\draw[-] (var2.east) -- (var3.west) node[scale =0.75, pos=0.5, above, align=left] {$2, 3$};
\draw[-] (var3.east) -- (var4.west) node[scale =0.75, pos=0.5, above, align=left] {$3, 4$};
\end{tikzpicture}

\hspace{1.5cm}

\begin{tikzpicture}[
    roundnode/.style={circle, draw=green!60, fill=green!5, very thick,
      minimum size=7mm},
]
\node[roundnode, scale = .75]      (var1)                              {$1, 2$};
\node[roundnode, scale = .75]        (var2)       [right=of var1] {$2, 3$};
\node[roundnode, scale = .75]      (var3)       [right=of var2] {$3, 4$};
\draw[-] (var1.east) -- (var2.west) node[scale=0.5, pos=0.5, above, align=left]
{$1, 3 \vert 4$};
\draw[-] (var2.east) -- (var3.west) node[scale=0.5, pos=0.5, above, align=left]
{$2, 4 \vert 3$};
\end{tikzpicture}

\hspace{1.5cm}

\begin{tikzpicture}[
    roundnode/.style={circle, draw=green!60, fill=green!5, very thick,
      minimum size=7mm},
]
\node[roundnode, scale = .75]      (var1)                 {$1, 3 \vert 2$};
\node[roundnode, scale = .75]        (var2)   [right=of var1] {$2, 4 \vert 3$};
\draw[-] (var1.east) -- (var2.west) node[scale=0.5, pos=0.5, above, align=left]
{$1, 4 \vert 2, 3$};
\end{tikzpicture}

\begin{tikzpicture}[
    roundnode/.style={circle, draw=green!60, fill=green!5, very thick,
      minimum size=7mm},
]
\node[roundnode, scale = .75]      (var1)                 {$1, 4 \vert 2, 3$};
\end{tikzpicture}

\caption{Illustration of a D-vine structure on 4 variables.
  \label{fig:fourdvine5}}
\end{figure}

A D-vine on $d$ variables is a collection of $d-1$ connected trees, number $t$
of which has $d-(t-1)$ nodes, and $d-t$ edges. The nodes of the first tree are
the $d$ original variables, whereas the nodes of the subsequent trees are the
edges from the previous tree. This is in general true for any vine, but for the
D-vine each tree has the same structure, namely that they are a single chain,
where each node has either one or two neighbours. A graphical illustration of
a D-vine is shown in Figure \ref{fig:fourdvine5}.

In the first tree, the arguments of the pair-copulas are the univariate margins,
whereas in the subsequent trees, they are conditional distributions. For
instance, $C_{i,j|v}$, with $i\neq j$, $(x_{i},x_{j})\notin \mathbf{x}_v$, is the
bivariate copula of $(X_{i}, X_{j})|\mathbf{X}_{v}$, which has the arguments
$F_{i|v}(x_{i}|x_{v})$ and $F_{j|v}(x_{j}|x_{v})$. The variables $(X_{i}, X_{j})$
are denoted the conditioned set and the variables $\mathbf{X}_{v}$ the
conditioning set.

These conditional distributions may be computed as follows
\citep{joe1996families,panagiotelis2012}:
\begin{equation}
  F_{i|j,v}(x_{i}|x_{j},\mathbf{x}_{v}) = \frac{\partial C_{i,j|v}
	(F_{i|v}(x_{i}|\mathbf{x}_{v}), F_{j|v}(x_{j}|\mathbf{x}_{v}))}
  {\partial F_{j|v}(x_{j}|\mathbf{x}_{v})},
  \label{eqn:cond_cont}
\end{equation}
when $X_{j}$ is continuous, and
\begin{equation}
  F_{i|j,v}(x_{i}|x_{j},\mathbf{x}_{v}) =
  \frac{C_{i,j\vert v}(F_{i\vert v}(x_{i}\vert\mathbf{x}_{v}),
    F_{j\vert v}(x_{j}\vert\mathbf{x}_{v})) -
    C_{i,j|v}(F_{i|v}(x_{i}|\mathbf{x}_{v}), F_{j|v}(x_{j}^{-}|\mathbf{x}_{v}))}
  {F_{j\vert v}(x_{j}\vert\mathbf{x}_{v}) -
    F_{j\vert v}(x_{j}^{-}\vert\mathbf{x}_{v})}
  \label{eqn:cond_discr},
\end{equation}
when $X_{j}$ is discrete, where $x_{j}^{-}$ is the largest value such 
that $F_{j}(x_{j}^{-}) < F_{j}(x_{j})$. In R-vines, and therefore in D-vines, the
copulas necessary to compute the conditional distributions, that are pair-copula
arguments, are always present in the previous tree.

\section{Estimation and model selection}\label{sec:estimation}
Fitting the model \eqref{eqn:model} consists of finding the
parameters that maximise the Bernoulli log-likelihood function
\begin{align}
	\mathcal{L} &= \sum_{i=1}^n
	\log\left({Pr(Y=1\vert \mathbf{X} = \mathbf{x}_i)}^{y_i}
	{Pr(Y=0\vert \mathbf{X} = \mathbf{x}_i)}^{1 - y_i}\right)\nonumber\\
	&= \sum_{i=1}^ny_ih(\mathbf{x}_i) -
   \sum_{i=1}^n\log\left(1 + \exp(h(\mathbf{x}_i))\right).
\label{eqn:loglik}   
\end{align}
For multiple reasons, it is computationally infeasible to solve this problem directly, due to the fact that the $h_m$-s come from different families of functions, and as discussed later, they will in this setting depend on subsets $\mathbf{x}_{S, m}$ of the covariates. Therefore, the number of possible combinations of functions and subsets of covariates for the components $h_m$ will quickly increase with both the number of covariates in the data and the number of components $M$ in the model.

To be able to fit this model within the restrictions, a stagewise approach to the fitting is taken, where one additional component of
the model is fitted in each `stage' of the process. That is, the $k$-th
is fitted by (approximately) maximising,

{\small
\begin{align}
  \mathcal{L}_{k} = \sum_{i=1}y_i
  \left(h_0 + \sum_{m=1}^{k-1}h_m(\mathbf{x}_i)
  + h_k(\mathbf{x}_i)\right) - 
  \sum_{i=1}\log\left(1 +
  \exp\left(h_0 + \sum_{m=1}^{k-1}h_m(\mathbf{x}_i)
  + h_k(\mathbf{x}_i)\right)\right),\label{eq:likelihood}
\end{align}
}%
\noindent whilst keeping $h_0 + \sum_{m=1}^{k-1}h_m(\mathbf{x}_i)$ fixed. This way of greedily estimating an additive model is a variation of
\textit{gradient boosting} \citep{gradientboosting}.

First, $h_{0}$ is estimated as
\[
h_{0} = \log\left(\frac{\widehat{Pr}(Y=1)}{\widehat{Pr}(Y=0)}\right) = \log\left(\frac{\bar{y}}{1-\bar{y}}\right),
\]
which is the empirical, unconditional log-odds. Then, at each stage $k$, for
$k=1,\ldots,M$, $h_k$ is fitted by regression, where the outcome is
(proportional to) the first order derivative of the likelihood with respect to
$h_k(\mathbf{x}),$ around $h_0 + \sum_{m=1}^{k-1}h(\mathbf{x}).$ In addition,
it is often common to scale the fitted function by a parameter $\gamma$, which is
often referred to as the \textit{learning rate}, in order to avoid over-fitting.
That is, $h_k(\mathbf{x})$ estimated as the regression function
\begin{align}
  h_k(\mathbf{x}) = \gamma\cdot m(\mathbf{x}) = 
  \gamma\cdot\mathbb{E}(\tilde{Y}^{[k]}\vert\mathbf{X}=\mathbf{x}),
\label{eqn:model_comp}
\end{align}
with
\begin{align}
\begin{split}  
  \tilde{y}_i^{[k]}
  &=y_i - \frac{\exp\left(h_0 + \sum_{m=1}^{k-1}h_m(\mathbf{x}_i)\right)}
  {1 + \exp\left(h_0 + \sum_{m=1}^{k-1}h_m(\mathbf{x}_i)\right)}, i=1,\ldots,n,\ k=1,\ldots,M,
\end{split}
\label{eqn:y_tilde}    
\end{align}
such that 
$\tilde{y}_i^{[k]} \propto \partial\mathcal{L}/\partial h_k(\mathbf{x}_i)\vert_{h_k(\mathbf{x}_i) = 0}$.

In this paper, the strategy is to select a subset $\mathbf{X}_{S,k}$ of the covariates, as will be outlined in Section \ref{subsec:cov_select}, and then think of  $\{(\tilde{Y}^{[k]}_i,\mathbf{X}_{S,k,i})\}_{i=1}^n$ as a random vector that one fits with a joint model $\mathcal{M}_k$, as described in Section \ref{subsec:mod_comp}. Subsequently, the $k$th component is computed as the conditional expectation
$h_k(\mathbf{x}) =
\gamma\mathbb{E}_{\mathcal{M}_{k}}
(\tilde{Y}^{[k]}\vert\mathbf{X}=\mathbf{x}) =
\gamma\mathbb{E}_{\mathcal{M}_{k}}(\tilde{Y}^{[k]}\vert\mathbf{X}_{S,k} =
\mathbf{x}_{S,k})$
under the fitted model $\mathcal{M}_{k}$, 
as explained in Section  \ref{subsec:mod_eval}.

\subsection{Choosing covariates, and fitting models}\label{subsec:cov_select}\label{subsec:mod_comp}

The computational complexity of both fitting and evaluating the components of the model grows quickly with their dimension. Therefore, each component $h_{k}(\mathbf{x})$ should depend only on a rather small subset $\mathbf{x}_{S,k}$ of the covariates, of a prespecified size $L$. This subset must be chosen for each component, for which the following greedy procedure is used. First, find the covariate $x_{k,1}$ for which
$h_{k}(\mathbf{x}) = \mathbb{E}(\tilde{Y}^{[k]}|x_{k,1})$ gives the largest
increase of the log-likelihood function $\mathcal{L}$ in \eqref{eqn:loglik}
with $h(\mathbf{x}) = h_{0} + \sum_{m=1}^{k}h_{m}(\mathbf{x})$, keeping
$h_{m}(\mathbf{x})$, $m=1,\ldots,k-1$, fixed from earlier iterations. Then, select the covariate $x_{k,2}$, such that
$h_{k}(\mathbf{x}) = \mathbb{E}(\tilde{Y}^{[k]}|x_{k,1},x_{k,2})$ increases
$\mathcal{L}$ the most. This is repeated until $\mathbf{x}_{S,k}$ consists of
$L$ covariates, as in that $\mathbf{x}_{S,k}=(x_{k,1},\ldots,x_{k,L})$.

In order to make the selection procedure more computationally efficient, a specialised algorithm based on some approximations and
simplifications is developed. Recall that the joint model of
$(\tilde{Y}^{[k]},\mathbf{X}_{S,k})$, on which $h_{k}(\mathbf{x})$ is built, is a vine. The structure of these vines is such that the conditional distribution $F_{\tilde{Y}^{[k]}|\mathbf{x}_{S,k}}(y|\mathbf{x}_{S,k})$, that
is needed in order to compute $\mathbb{E}(\tilde{Y}^{[k]}|\mathbf{x}_{S,k})$,
is obtained with a simple formula (\eqref{eqn:cond_cont} or \eqref{eqn:cond_discr})
if $\tilde{Y}^{[k]}$ is in the conditioned set of the top node, but may
require multidimensional integration otherwise. Therefore, the vines are constructed such that this is the case, as is also suggested by \citet{chang2019prediction}. Further, when adding one more
covariate to $\mathbf{x}_{S,k}$, and thus to the vine, the structure of the part of the model that was present prior to adding the covariate should stay the same. An easy way to obtain both these requirements, is to let the vine be a D-vine with the order
$\tilde{Y}^{[k]},X_{k,1},\ldots,X_{k,L}$ in the first tree. An illustration of such a vine structure with three covariates is shown in Figure \ref{fig:fourdvine3}.

\begin{figure}
  \centering
\begin{tikzpicture}[
    roundnode/.style={circle, draw=green!60, fill=green!5, very thick,
      minimum size=7mm},
]
\node[roundnode]      (var1)                              {$Y^{[k]}$};
\node[roundnode]        (var2)       [right=of var1] {$j_1$};
\node[roundnode]      (var3)       [right=of var2] {$j_2$};
\node[roundnode]        (var4)       [right=of var3] {$j_3$};

\draw[-] (var1.east) -- (var2.west) node[scale =0.75, pos=0.5, above, align=left] {$Y^{[k]}, j_1\,$};
\draw[-] (var2.east) -- (var3.west) node[scale =0.75, pos=0.5, above, align=left] {$j_1, j_2$};
\draw[-] (var3.east) -- (var4.west) node[scale =0.75, pos=0.5, above, align=left] {$j_2, j_3$};
\end{tikzpicture}

\hspace{1.5cm}

\begin{tikzpicture}[
    roundnode/.style={circle, draw=green!60, fill=green!5, very thick,
      minimum size=7mm},
]
\node[roundnode, scale = .75]      (var1)                              {$Y^{[k]}, j_1$};
\node[roundnode, scale = .75]        (var2)       [right=of var1] {$j_1, j_2$};
\node[roundnode, scale = .75]      (var3)       [right=of var2] {$j_2, j_3$};
\draw[-] (var1.east) -- (var2.west) node[scale=0.5, pos=0.5, above, align=left]
{$Y^{[k]}, j_2 \vert j_3$};
\draw[-] (var2.east) -- (var3.west) node[scale=0.5, pos=0.5, above, align=left]
{$j_1, j_3 \vert j_2$};
\end{tikzpicture}

\hspace{1.5cm}

\begin{tikzpicture}[
    roundnode/.style={circle, draw=green!60, fill=green!5, very thick,
      minimum size=7mm},
]
\node[roundnode, scale = .75]      (var1)                 {$Y^{[k]}, j_2 \vert j_1$};
\node[roundnode, scale = .75]        (var2)   [right=of var1] {$j_1, j_3 \vert j_2$};
\draw[-] (var1.east) -- (var2.west) node[scale=0.5, pos=0.5, above, align=left]
{$Y^{[k]}, j_3 \vert j_1, j_2$};
\end{tikzpicture}

\begin{tikzpicture}[
    roundnode/.style={circle, draw=green!60, fill=green!5, very thick,
      minimum size=7mm},
]
\node[roundnode, scale = .75]      (var1)                 {$Y^{[k]}, j_3 \vert j_1, j_2$};
\end{tikzpicture}

\caption{Illustration of a possible D-vine structure with 3 covariates.
  \label{fig:fourdvine3}}
\end{figure}

Further, when both $\tilde{Y}^{[k]}$ and all the covariates are continuous, we have that for $j=1,\ldots,L$
\begin{align*}
\mathbb{E}\left(\tilde{Y}^{[k]} \,\vert\, x_{k,1},\ldots,x_{k,j} \right) &=
  \int_{-1}^1y\frac{\partial F(y\vert x_{k,1},\ldots,x_{k,j})}{\partial y}dy\\
  &=\int_{0}^1F^{-1}_{Y^{[k]}}(u)
  c_{Y^{[k]},X_{k,j}\vert x_{k,1},\ldots,x_{k,j-1}}(u, F(x_{k,j} \vert x_{k,1},
  \ldots,x_{k,j-1}))du,
\end{align*}
where $c_{Y^{[k]},X_{k,j}\vert x_{k,1},\ldots,x_{k,j-1}}$ is the density of the
copula $C_{Y^{[k]},X_{k,j}\vert x_{k,1},\ldots,x_{k,j-1}}$ of\\
$(Y^{[k]},X_{k,j})\vert x_{k,1},\ldots,x_{k,j-1}$ (or simply the unconditional
copula of $(Y^{[k]},X_{k,1})$ for $j=1$). In Section \ref{subsec:mod_eval}, a procedure for computing this conditional mean for a general copula $C$ is described, also when (some of) the covariates are discrete. That procedure is however too slow when it must be repeated as many times as in the covariate selection.
Therefore, the following approximations are useful to make. Let all the copulas in the vine be Gaussian, and approximate the quantile function of $Y^{[k]}$ by
$$
F^{-1}_{Y^{[k]}}(u) \approx \gamma_0  +
\sum_{j=1}^m \gamma_j\Phi^{-1}(u)^j,
$$
where $\Phi$ is the cdf of the standard normal distribution, and $m=2$ typically gives a sufficiently good approximation for this application. Writing
$$u_{j\vert 1\ldots j-1}=F(x_{k,j} \vert x_{k,1},\ldots,x_{k,j-1}),$$ the conditional
expectation now simplifies to
\begin{footnotesize}
\begin{align*}
\mathbb{E}\left(\tilde{Y}^{[k]} \,\vert\, x_{k,1},\ldots,x_{k,j} \right)
&\approx \int_{0}^1\left(\gamma_0  +
\sum_{j=1}^m \gamma_j\Phi^{-1}(u)^j\right)
\frac{1}{\sqrt{1-\rho^{2}}}\phi\left(\frac{\Phi^{-1}(u) - \rho\Phi^{-1}
\left(u_{j\vert 1\ldots j-1}\right)}{\sqrt{1-\rho^{2}}}\right)
\phi\left(\Phi^{-1}(u)\right)du\\
&=\gamma_0 + \sum_{j=1}^m\gamma_j\int_{-\infty}^\infty
z^j\frac{1}{\sqrt{1-\rho^{2}}}\phi\left(\frac{z - \rho\Phi^{-1}
\left(u_{j\vert 1\ldots j-1}\right)}{\sqrt{1-\rho^{2}}}\right)dz\\ &= \gamma_0 + \sum_{j=1}^m\gamma_j\mathbb{E}(Z^j),
\end{align*}
\end{footnotesize}
where $\phi$ is the pdf of the standard normal distribution, $\rho$ is the
parameter of the copula\\
$C_{Y^{[k]},X_{k,j}\vert x_{k,1},\ldots,x_{k,j-1}},$ and $Z$
follows a normal distribution with mean
$$\rho\Phi^{-1}\left(F(x_{k,j} \vert x_{k,1},\ldots,x_{k,j-1})\right),$$ and standard
deviation $\sqrt{1-\rho^{2}}$. The weights $\gamma_0, \gamma_1, \dots, \gamma_m$
are found by solving the problem
\begin{small}
$$
\underset{\gamma_0, \gamma_1, \dots, \gamma_m}{\text{argmin}}
\sum_{i=1}^n\left(\tilde{y}_{i}^{[k]} - \gamma_0  -
\sum_{j=1}^m \gamma_j\Phi^{-1}(\hat{u}_i^*)^j\right)^2, \quad \mbox{where} \quad \hat{u}_i^* = \hat F\left(Y^{[k]}\right) =
\frac{1}{n}\sum_{i=1}^n\mathcal{I}_{[\tilde{y}_{i}^{[k]},\infty)}(Y^{[k]}),
$$
\end{small}
where $\tilde{y}_{i}^{[k]}$, $i=1,\ldots,n$, are
the outcomes used in the $k$th iteration, defined in \eqref{eqn:y_tilde}, and $\mathcal{I}_A(a)$ is the indicator function, which is $1$ when $a \in A$, and $0$ otherwise.
Since $Y^{[k]} = Y - \hat{p}^{[k-1]} \in (-1, 1)$, and the approximated quantile
function can take all values in $\mathbb{R}$, the estimate of 
$
h_k(\mathbf{x}) = \mathbb{E}\left(\tilde{Y}^{[k]} \,\vert\, \mathbf{x}_S\right)
$
is truncated to 
$
\hat h_k(\mathbf{x}) = \text{sign}(\tilde{h}_k(\mathbf{x}))
\text{min}\left\{\vert\tilde{h}_k(\mathbf{x})\vert, 1\right\},
$
where $\tilde{h}_k$ is the untruncated estimate. Further, some of the covariates
are not in fact continuous, as assumed in the above computations. This is solved by adding a small Gaussian noise term to each discrete covariate, which is often referred to as \textit{jittering} \citep{nagler2018generic}. As jittering may
introduce bias in the estimates \citep{nagler2018generic}, this should only be done in the covariate selection procedure in order to increase the computational efficiency, but not when fitting the models. Further, when $Y^{[k]}$ is discrete, which typically is
only the case for $k = 1$, the above conditional expectation is instead computed
as
\begin{align*}
\mathbb{E}\left(\tilde{Y}^{[k]} \,\vert\, x_{k,1},\ldots,x_{k,j} \right) &=
\sum_{y \in \Omega_{\tilde{y}}}y \left(F(y\vert x_{k,1},\ldots,x_{k,j})-
F(y^{-}\vert x_{k,1},\ldots,x_{k,j})\right),
\end{align*}
where $\Omega_{\tilde{y}}$ is the sample space of $Y^{[k]},$ and
$$F(y\vert x_{k,1},\ldots,x_{k,j}) =
\Phi\left(\left(z - \rho\Phi^{-1}\left(u_{j\vert 1\ldots j-1}\right)\right)/
\sqrt{1-\rho^{2}}\right).$$

After $L$ covariates are selected using this greedy approach, a vine is fitted to $(\tilde{Y}^{[k]},X_{k,1},\ldots,$ $X_{k,L})$. The structure of this vine is kept as the D-vine from the covariate selection. However, the copulas are no longer
restricted to being Gaussian. This increases the flexibility of the model, and
may increase the prediction performance of the model. Further, the discrete
covariates are now treated as such. Moreover, the parametric pair-copulas are selected and
fitted one by one, as proposed in \cite{dissmann2013selecting}, but with a prespecified vine structure. The copula parameters are fitted in a sequential, two-stage maximum likelihood procedure
(\citep{joe1996families,ko2019}), and the selected parametric copula model is the one among a list of candidates with the highest value of AIC.
\subsection{Evaluating  the model components}\label{subsec:mod_eval}
As mentioned earlier, one needs to be able to compute the
conditional expectation $\mathbb{E}(\tilde{Y} \vert \mathbf{X}=\mathbf{x}),$
given a pair-copula joint model for $(\tilde{Y}, \mathbf{X}),$ and a point
$\mathbf{x},$ to evaluate at. Depending on whether $\tilde{Y}$ is a discrete or
a continuous variable, this problem is either quite simple, or somewhat more
complicated. In the discrete case, the expectation can be computed as follows
\begin{align*}
	\mathbb{E}(\tilde{Y} \vert \mathbf{X} = \mathbf{x})
	&= \sum_{\tilde{y} \in \Omega_{\tilde{Y}}}
	\tilde{y}Pr(\tilde{Y}=\tilde{y} \vert \mathbf{X}=\mathbf{x})
	= \sum_{\tilde{y} \in \Omega_{\tilde{Y}}} \tilde{y}
	\left(F(\tilde{y}\vert\mathbf{x}) - F(\tilde{y}^{-}\vert\mathbf{x})\right),
\end{align*}
where $\Omega_{\tilde{Y}}$ denotes the sample space of $\tilde{Y}$, and
$F(\tilde{y} \vert \mathbf{x})$ can be computed directly from the copulas in the pair-copula construction, as a consequence of how the model is selected (see Section \ref{subsec:mod_comp}).

In the case where $\tilde{Y}$ is continuous, the conditional expectation is an integral that cannot be computed analytically. Instead, one may use
either Monte Carlo methods or some type of numerical integration scheme, such
as the trapezoidal rule. The former is in this case much slower than the latter
when requiring the same precision. Further, numerical integration requires
evaluation of the conditional density of $\tilde{Y}$, which is undesirable, as the evaluation of the conditional density is quite time consuming.
Instead, an approximation based on the following idea can be used.

Assume that computing the integral
$\int_a^b\tilde{y}f(\tilde{y}\vert \mathbf{x})d\tilde{y}$ is of interest.
In contrast to the trapezoidal rule, in which one would approximate
$\tilde{y}f(\tilde{y}\vert \mathbf{x})$ by the secant that interpolates
$\tilde{y}f(\tilde{y}\vert \mathbf{x})$ between $a$ and $b$, so that the
integral becomes $(b-a)\frac{bf(b) - af(a)}{2}$, we replace the integrand by
$y^{*}(a, b) f(\tilde{y}\vert \mathbf{x}),$ where $y^{*}(a, b)$ is a constant.
The integral over this function is
\begin{align*}
	\int_{a}^b{y}^{*}(a, b) f(\tilde{y}\vert \mathbf{x}) d\tilde{y} = {y}^{*}(a,b)
	\left(F(b\vert \mathbf{x}) - F(a \vert \mathbf{x})\right).
\end{align*}
A motivation for this idea is that for a set of disjoint intervals
$\{A_l\}_{l=1}^K$, such that 
$$
P(\tilde{Y} \in \cup_{l=1}^KA_l) = 1,
$$
we have that
\begin{align*}
	\mathbb{E}(\tilde{Y} \vert \mathbf{X}=\mathbf{x}) =
	\sum_{l}\mathbb{E}(\tilde{Y} \vert \tilde{Y} \in A_l, \mathbf{X}=\mathbf{x})
	P(\tilde{Y}\in A_l\vert \mathbf{X} = \mathbf{x}).
\end{align*}
If the interval $A_l$ is narrow, then the value of
$\mathbb{E}(\tilde{Y} \vert \tilde{Y} \in A_l, \mathbf{X}=\mathbf{x})$
cannot vary much, and therefore it seems reasonable to replace it by a constant,
$y^{*} \in A_l$. A natural choice is to let the constant ${y}^{*}$ be an estimate
of the expectation of the random variable $\tilde{Y}$ truncated to the interval
$[a, b]$, without conditioning on $\mathbf{X}$. For convenience, one can let ${y}^{*}$ be an estimate of the median of
$\tilde{Y}\cdot\mathcal{I}_{[a, b]}(\tilde{Y})$. The local median and local expectation should not be too far apart, especially when $a$ and $b$ are close, such that there are few observations of $\tilde{Y}$ in this interval. Therefore, using the median instead of the expectation should not have a big impact on the approximation.

Using this idea, the support $\Omega_{\tilde{Y}}$ of $\tilde{Y}$ can be
divided into $K$ subintervals $\{(y_{l-1}, y_l)\}_{l=1}^K,$ where $y_0$ and $y_K$ correspond to the endpoints of the support of $\tilde{Y}$, which in this application is always an interval that is bounded in both directions. Let
$$y_l = F_{\tilde{Y}}^{-1}\left(\frac{l}{K}\right) \mbox{   and   }
{y}^{*}_l = F_{\tilde{Y}}^{-1}\left(\frac{l}{K} + \frac{1}{2K}\right),$$
where $F_{\tilde{Y}}^{-1}$ is estimated by the empirical quantile function
$\hat{F}_{\tilde{Y}}^{-1}(u)=\tilde{y}_{\lceil un \rceil}$, $\lceil y \rceil$ being the smallest
integer larger than $y$ and $\tilde{y}_{(1)}\leq\ldots\leq\tilde{y}_{(n)}$,
given in \eqref{eqn:y_tilde}, are sorted in ascending order. The expectation can then be estimated as
\begin{align*}
	\int_{\Omega_{\tilde{Y}}}\tilde{y}f(\tilde{y}\vert\mathbf{x})d\tilde{y} &=
	\sum_{k=1}^K\int_{y_{k-1}}^{y_k}\tilde{y}f(\tilde{y}\vert \mathbf{x})d\tilde{y} \approx \sum_{k=1}^K{y}^{*}_k\left(F(y_k\vert \mathbf{x}) -
	F(y_{k-1}\vert \mathbf{x})\right).
\end{align*}

\subsection{Modelling the marginal distributions}\label{subsec:marg_model}

\begin{figure}
  \centering
  \includegraphics[width=300pt]{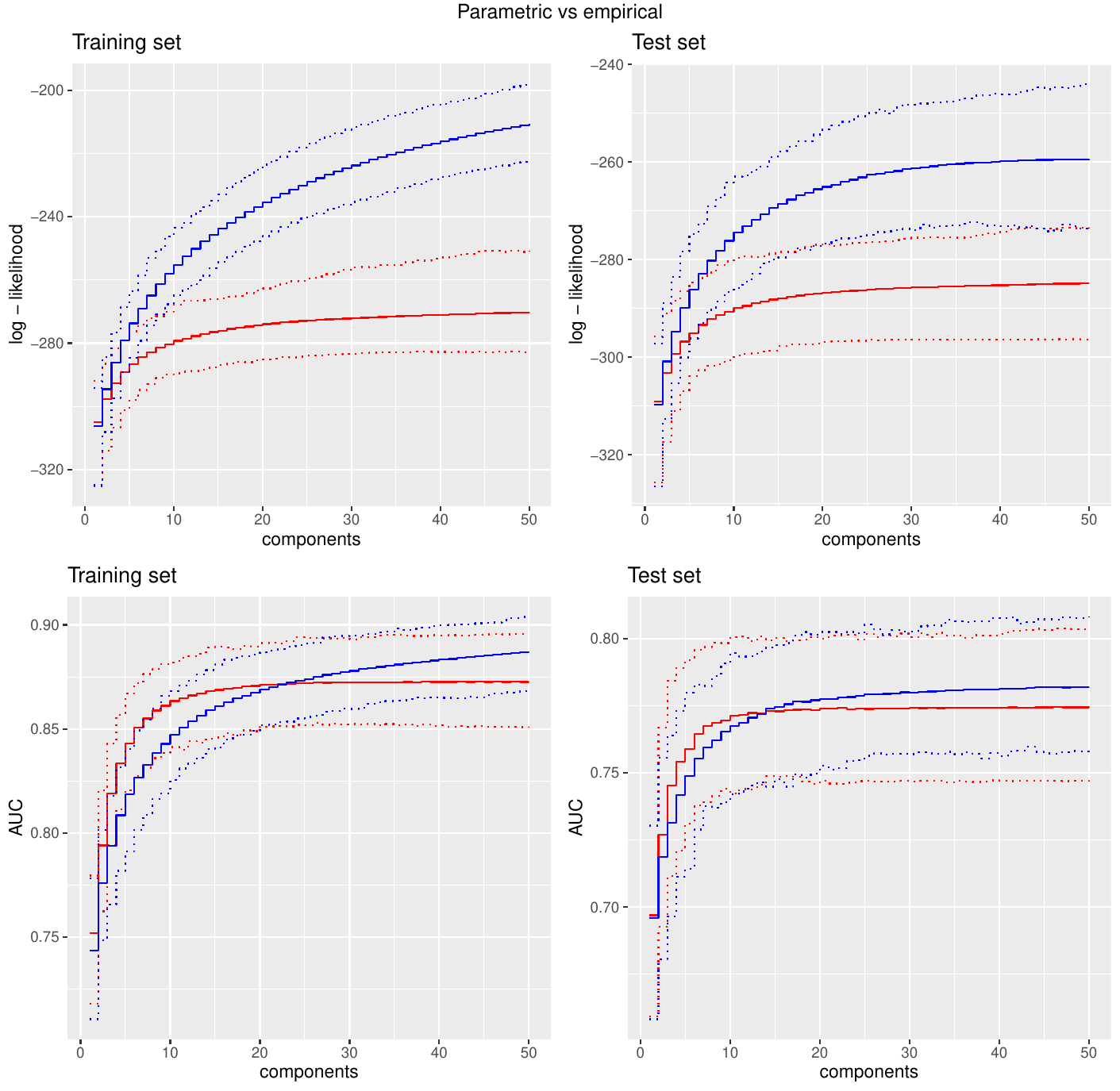}
  \caption{Averaged results of 100 simulations comparing empirical margins to
    parametric margins, represented as red and blue solid lines, respectively.
    The corresponding dotted lines show the 10th and 90th percentiles.
    \label{fig:par_v_emp}}
\end{figure}

Up until this point, no mention has been made of the modelling of the marginal distributions of $X,$ $Y,$ and each $\tilde{Y}^{[k]}.$ A common approach in copula applications is to model the marginal distributions with the nonparametric estimator
\begin{align*}
\hat{F}(x) = \begin{cases}
  \frac{1}{n + 1}\sum_{i=1}^n\mathcal{I}_{[x_i, \infty)}(x) \text{ if } x \text{ is continuous}\\
    \frac{1}{n}\sum_{i=1}^n\mathcal{I}_{[x_i, \infty)}(x) \text{ if } x \text{ is discrete}\\
  \end{cases},
\end{align*}
where the factor $\frac{1}{n+1},$ as opposed to $\frac{1}{n},$ is used to
avoid numerical problems for continuous variables. The alternative to using this
nonparametric estimator is to model the marginal distributions using a parametric
model for each margin. The nonparametric estimator may be better suited in a
conventional setting when fitting copula distributions to data where the main
interest is inference about the parameters of the copula, and possibly the
pair-copula structure. 
It is not given that this is the case in an application
such as this. Therefore, it is interesting to make a comparison between parametric and nonparametric estimators of the margins for the 
modelling approach outlined here.

For the parametric approach, all the continuous margins are modelled
as Gaussian random variables, all dichotomous variables as Bernoulli random variables, and ordinal
random variables with domain $\{a_j\}_{j=1}^J$, that is a collection of points in
$\mathbb{R}$, with a distribution with $J$ parameters ${p_j}_{j=1}^J,$ where
$p_j = P(X = a_j).$ If there are any discrete variables without a natural
ordering on the domain, these are instead re-coded as one or more Bernoulli random variables. In their applications the authors have not encountered any covariates that are counting variables, but this approach could easily be adapted by using, for instance, a Poisson model.

The performance of the two approaches for the margins was explored in a simulation example, where the simulation model is set up to mimic the proposed method. First, 100 data sets were drawn in the following way. A $(1000 \times 100)$ covariate matrix $\mathbf{X}$ was generated by a Gaussian copula with $p\times p$ correlation matrix $R$, and the following marginal
distributions; 25 were $Gamma(2,2)$, 25 were $Beta(3,1)$, 25 were Bernoulli with individual parameters, equidistantly placed on the interval $[0.2, 0.8]$, and the last 25 were t-distributed with individual degrees of freedom parameters, equidistantly placed on the interval $[3, 8]$. Then, $[\mathbf{X}_1, \dots, \mathbf{X}_{30}]$, which are $30$ different subvectors of
$\mathbf{X}$ were drawn with replacement, so that the same variables can appear in more than one subvector. Each of the subvectors contain $4$ of the columns of
$\mathbf{X}$. For each subvector $\mathbf{X}_m,$ an R-vine model was defined on $(S_m, \mathbf{X}_m),$ where $S_m$ has to be a leaf node in the first tree of the R-vine, so that 
$\mathbb{E}(S_m\vert \mathbf{X}_m)$, can be computed easily, using the method described in Section
\ref{subsec:mod_eval}. The marginal distribution of $S_m$ was specified to be uniform on the interval $(-1, 1).$ The predictor was then defined as the sum
$f(x) = \beta_0 + \sum_{m=1}^M\mathbb{E}(S_m\vert\mathbf{X}_m),$
which is transformed into probabilities via the link function so that
$$
P(Y=1 \vert \mathbf{x}) = \frac{\exp(\mathbf{x})}{1 + \exp(\mathbf{x})}.
$$
The intercept $\beta_0$ was adapted so that $P(Y=1) \approx \frac{1}{3}.$
Lastly, $Y$ was drawn from a Bernoulli distribution with these probabilities.

Each of these 100 datasets were divided into a training and a test set of equal
size. Then, models were fitted to the training set, using both the parametric and the nonparametric approach for the margins, and the performance evaluated on the test set. For all the models that were fitted, the learning rate was specified as $\gamma=1$, each model component was built from $L=4$ covariates, and $M=50$ such components
were used. The results of these simulations are summarised in Figure \ref{fig:par_v_emp},
in terms of the likelihood and the AUC, both for the training data and the test
data. One can see that the parametric approach performed the best in all cases, even though the parametric models we fit are incorrectly specified for all of the continuous variables. Based on this, we choose to use a parametric approach for the marginal distributions, for these types of models.
Gaussian mixture models were also considered for the margins. However, these lead to worse performance than simple Gaussian models.

\section{Simulation study}\label{sec:sim_study}
In order to explore different qualities and aspects of the modelling strategy, a simulation study was carried out. 
The design of the model suggests that it might have an advantage over simpler models in cases where the relationship between the response and the covariates is determined not only by main effects, but also interactions. However, a more complex model requires more data to achieve the same accuracy as a simpler model, so it is expected that this model will be outperformed by simpler and more robust models if there is little data available for estimation.
Therefore, the effect of varying the parameters of the model is investigated, specifically how the number of covariates included in each model component impacts the stability of the model in Section \ref{subsec:study1}. In addition, a comparison is made with alternative models in Section \ref{subsec:study2}. 

\subsection{Simulation models}
The simulation model described in Section \ref{subsec:marg_model} is used, as well as three additional model scenarios that are outlined below, simulating a binary response variable $Y$, as well as a $100$ dimensional covariate vector
$\mathbf{X}$. In settings one, three and four, the covariate vector was drawn as
specified in Section \ref{subsec:marg_model}. What separates the different
scenarios, is the conditional distribution of the outcome, given the covariates,
that is, the form that $Pr(Y=1|\mathbf{X}=\mathbf{x})$ takes.
\begin{itemize}
  \item{Setting one: Linear predictor with main effects and interaction terms.}\\
    The response $Y$ is drawn from the Bernoulli
    distribution with $p(x)$ determined by a linear predictor $\eta(x)$ that
    includes a selection of the covariates, with interaction terms, transformed
    via the link function $g(x) = \frac{\exp(x)}{1 + \exp(x)}.$

  \item{Setting two: joint R-vine model}\\
    Joint model for $(Y, \mathbf{X})$
    expressed as an R-vine model, where a subset $\mathbf{X}_{S}$ of $\mathbf{X}$ and $Y$ are dependent. In this way, $Y$ will depend on $\mathbf{X},$
    potentially with very high order interactions, depending on the dimension
    $d$ of $\mathbf{X}_{S}$. The R-vine structure is specified by drawing randomly from a uniform distribution of all R-vines on $d+1$ variables \citep{cooke2015}. Next, the pair-copulas are specified, which are set
    to be either Gaussian, Clayton, or Gumbel copulas. Then, the
    parameters for each of these are set, by first specifying the Kendall's $\tau$. Each Kendall's tau is drawn from a uniform distribution on
    $(0, b_t),\, t=1,\, ...,\, t+1,$ for each level $t$ of the R-vine, where
    $b_1 > b_2, ... > b_T$. Thus, the pairwise dependencies in the pair-copulas
    vary, and are strongest at the bottom of the R-vine structure, which is the
    most influential level for the overall dependence. Finally, the
    Kendall's $\tau$s are converted to corresponding parameter values for the pair-copulas.
    The marginal distributions of $\mathbf{X}$ are the same composition
    of distributions as in the three other simulation settings. The data
    $(y,\mathbf{x})$ are drawn as follows. First, $d+1$ dependent uniform
    numbers are generated from the R-vine, and $p-d$ independent uniforms are
    also generated. The first among the dependent uniforms is transformed to $y$
    with the quantile function of the Bernoulli distribution with parameter $p_{y}$.
    The remaining $p$ uniforms are transformed to $\mathbf{x}$ via the quantile
    functions of the margins specified for $\mathbf{X}$.

  \item{Setting three: additive copula effect model}\\
  The third simulation model is a model that is similar to the additive copula regression models that are outlined in this paper, and is described in detail in section \ref{subsec:marg_model}.

  \item{Setting four: linear predictor with interactions effects only}\\
  Only interaction terms with either 2, 3 or 4
  covariates are included, without any first order effects.

\end{itemize}

\subsection{Initial study: Performance when varying the model parameters}
\label{subsec:study1}
In the first simulation study, datasets are drawn from each of the three settings, models fitted with either 2, 3, 4, 5 or 6 covariates each, with a maximum of 600,
500, 400, 300 or 200 components, respectively. Further, the learning rate
$\lambda$ is varied between 1/3, 2/3 and 1. For each repetition of one
experiment, three data sets are drawn, a training, a validation and a test set, with $n = 500$ observations and $p = 100$ covariates, where the marginal
probability of $Y = 1$ is fixed to be $0.33$, by adjusting the intercept in each simulation
in Settings 1, 3 and 4, and by setting $p_{y}=0.33$ in setting 2. In all of the
simulations, 25 of the covariates are Bernoulli-distributed, 25 follow a
gamma-distribution, 25 a beta-distribution, and the remaining 25 follow a
t-distribution, in the same way as described in Section \ref{subsec:marg_model}.

In the first setting, the predictor consists of
20 univariate effects, 45 two-way interaction terms, 20 three-way interaction
terms, and one 4-way interaction terms. In the second, $Y$ depends on
$d=20$ of the covariates, through a 21-dimensional joint R-vine. In the third
setting the predictor consists of 30 components that each depend on four
covariates in the same way as described in section \ref{subsec:marg_model}, and
in the fourth, the predictor consists of 30 3-way interactions.

First, the training set is used to fit a model with $M$ components, for a given learning rate $\lambda$ and a given number of covariates in each
component. Subsequently, the validation set is used set to select number $M^{*} \leq  M$ of components to include in the model, which maximises either the likelihood or the AUC. The chosen model is then evaluated on the test set. 
The results of these simulations 
for the four settings are summarised in Table
8 in the supplementary material and Table 9 in the supplementary material.
They indicate that in general,
the model is able to achieve a better degree of separation for the predictions
when the number of covariates in each component is increased. The difference in AUC
is not that dramatic, though, and models with larger components require more time
to be fitted even if they need a smaller number $M^{*}$ of components to give a
good fit. One could therefore choose to fit components with for instance three
covariates, to get a model that gives rather good predictions, and takes shorter
time to fit compared to a model with larger components. An interesting observation
is that the variance of the AUC and the log-likelihood does not seem to increase
when larger components are used, except in the fourth setting, and in some cases
rather seems to decrease, which was not something that was expected, a priori.
In terms of the likelihood, it is not always the case that larger components give
a better fit, and the best choice of component size seems to correspond in Setting
one and four to the complexity of the simulation model, but not necessarily in
Setting three. The learning rate does not seem to affect the results substantially.
However, as illustrated by Table 10 in the supplementary material, the learning rate and the number of
covariates in each component both affect the number of components $M$ that were selected, in that
increasing either of them decreases the number of components that are selected.

\begin{table}[htbp]
  \centering
  \resizebox{\textwidth}{!}{\begin{minipage}{\textwidth}\label{tab:complik}
        \caption{Results of the simulations in terms of the likelihood for the comparison study, with the standard deviation in parentheses.}
        \begin{tabular}{c|c|c|c|c}
\hline
 Interaction size & Logic regression & copulaboost & Ridge & xgboost \\
\hline
\multicolumn{2}{c}{33\% discrete}\\
\hline
2 & -315.676 (14.516) & -298.657 (10.97) & -277.325 (12.366) & -291.161 (13.035) \\ 
3 & -320.802 (10.642) & -309.33 (7.902) & -294.859 (11.067) & -314.03 (10.172) \\ 
4 & -320.089 (9.95) & -302.749 (8.916) & -303.585 (8.391) & -314.715 (7.82) \\ 
\hline
\multicolumn{2}{c}{67\% discrete}\\
\hline
2 & -310.235 (15.198) & -286.595 (12.253) & -264.898 (11.818) & -293.971 (12.14) \\ 
3 & -322.24 (10.914) & -311.088 (7.359) & -298.804 (10.549) & -315.651 (8.667) \\ 
4 & -324.167 (9.725) & -316.793 (6.839) & -313.77 (7.495) & -325.991 (7.159) \\ 
\hline
\multicolumn{2}{c}{100\% discrete}\\
\hline
2 & -303.442 (15.352) & -284.191 (10.258) & -264.966 (9.909) & -288.824 (11.547) \\ 
3 & -318.91 (14.033) & -310.438 (7.478) & -296.857 (10.006) & -310.459 (10.112) \\ 
4 & -322.953 (9.755) & -316.275 (7.168) & -313.085 (7.97) & -324.817 (7.443) \\ 
\hline
\end{tabular}
      \end{minipage}}
\end{table}

\begin{table}[htbp]
  \centering
  \resizebox{\textwidth}{!}{\begin{minipage}{\textwidth}\label{tab:compauc}
        \caption{Results of the simulations in terms of the AUC for the comparison study, with the standard deviation in parentheses.}
        \begin{tabular}{c|c|c|c|c}
\hline
 Interaction size & Logic regression & copulaboost & Ridge & xgboost \\
\hline
\multicolumn{2}{c}{33\% discrete}\\
\hline
2 & 0.641 (0.029) & 0.741 (0.027) & 0.74 (0.025) & 0.728 (0.027) \\ 
3 & 0.595 (0.03) & 0.685 (0.03) & 0.686 (0.031) & 0.641 (0.035) \\ 
4 & 0.571 (0.032) & 0.656 (0.031) & 0.647 (0.031) & 0.618 (0.028) \\ 
\hline
\multicolumn{2}{c}{67\% discrete}\\
\hline
2 & 0.664 (0.031) & 0.772 (0.025) & 0.769 (0.025) & 0.736 (0.024) \\ 
3 & 0.596 (0.035) & 0.663 (0.03) & 0.667 (0.027) & 0.626 (0.03) \\ 
4 & 0.539 (0.028) & 0.576 (0.031) & 0.586 (0.029) & 0.549 (0.029) \\ 
\hline
\multicolumn{2}{c}{100\% discrete}\\
\hline
2 & 0.688 (0.034) & 0.772 (0.023) & 0.771 (0.022) & 0.753 (0.024) \\ 
3 & 0.617 (0.032) & 0.67 (0.029) & 0.675 (0.028) & 0.659 (0.029) \\ 
4 & 0.547 (0.032) & 0.577 (0.028) & 0.587 (0.028) & 0.57 (0.03) \\ 
\hline
\end{tabular}
      \end{minipage}}
\end{table}

\begin{table}[htbp]
  \centering
  \resizebox{\textwidth}{!}{\begin{minipage}{\textwidth}\label{tab:complik_2}
        \caption{Results of the simulations in terms of the likelihood for the comparison based on data from 'Setting 2', with the standard deviation in parentheses.}
        \begin{tabular}{c|c|c|c|c}
\hline
No. of covariates & copulaboost & xgboost  & Logic regression & Ridge\\
\hline
2 & -304.130 (18.975) & -656.057 (7.129) & -664.428 (7.855) & -352.858 (17.087)\\ 
3 & -291.413 (17.233) & -656.146  (7.058)  & -664.428 (7.855) & -352.858 (17.087)\\ 
4 & -288.461 (17.034) & -655.965 (7.112) & -664.428 (7.855) & -352.858 (17.087)\\ 
\hline
\end{tabular}
      \end{minipage}}
\end{table}

\begin{table}[htbp]
  \centering
  \resizebox{\textwidth}{!}{\begin{minipage}{\textwidth}\label{tab:compauc_2}
        \caption{Results of the simulations in terms of the AUC for the comparison based on data from 'Setting 2', with the standard deviation in parentheses.}
        \begin{tabular}{c|c|c|c|c}
\hline
No. of covariates & copulaboost & xgboost  & Logic regression & Ridge\\
\hline
2 &  0.932 (0.010) & 0.904 (0.011) & 0.865 (0.016) & 0.915 (0.010)\\ 
3 &  0.939 (0.008) & 0.904 (0.011) & 0.865 (0.016) & 0.915 (0.010)\\ 
4 & 0.940 (0.008) & 0.904 (0.011) & 0.865 (0.016) & 0.915 (0.010)\\ 
\hline
\end{tabular}  
      \end{minipage}}
\end{table}

\subsection{Comparison to other methods}
\label{subsec:study2}
A comparison study of the method was conducted, against other natural alternatives, such as logic regression and an additive tree model, here fitted using xgboost \citep{chen2016xgboost}. Ridge-penalised logistic regression without interaction effects was also
included as a reference. First, data from logistic regression models
was simulated, with exclusively interaction terms of order 2, 3, or 4, that is, the type of model termed 'Setting four' in the previous section. In
addition to varying the number of covariates included in the interaction terms, the number of discrete (0-1) covariates in the data sets was also varied. Whereas in the first study the proportion of discrete covariates was kept at 25\%, this is now either 33\%, 67\% or 100 \%. The rest of the covariates are divided into in 3 groups of equal size, which follow a gamma, beta or t-distribution respectively.

Each model component was built from 3 covariates, and the
learning rate was set to $\frac{1}{3}$. The number $M$ of components was chosen
as the one that gave the highest likelihood value on a validation set of the
same size as the training set. For the additive tree model(xgboost), the learning rate was fixed at the default value, and the number of model components was chosen
based on 10-fold cross validation, using the AUC as a measure. For the logic regression model, all the continuous covariates had to be dichotomized, which was done by replacing the original variables $x_{i}$ by indicators of whether
the values $x_{ij}$ were larger than the median of $x_{i}$ or not. The model
was selected by fitting models with 3 to 5 trees and a total number of leaves
between 4 and 128 (5 trees and 128 leaves are the maximum values) and selecting
the model with the best score. For the ridge-penalised logistic regression,
the penalisation parameter was chosen by 10-fold cross validation, using the
deviance as a measure.

The results of these simulations are summed up in terms of the
log-likelihood and the AUC based on independent test sets in Tables
\ref{tab:complik} and \ref{tab:compauc}, respectively. The main purpose
here is to assess how the performance of our method is affected by having many
discrete variables in the data set, as this should have an
adverse affect, compared to how this affects other methods, such as logic
regression and tree based models, that are designed with discrete data in mind
or especially well suited to discrete data.

Judging from the results in the two mentioned tables, the performance is
less adversely affected by an increase in the number of discrete
covariates, compared to the other models, than expected. In terms of
the absolute numbers, the additive copula regression model was better on average than logic regression and xgboost in every setting, judged both by the likelihood and the AUC. Compared to the penalised logistic regression model, the model performed better in some settings and worse in others, as judged by the AUC. Further, likelihood was lower than the logistic regression model, for all but one case, but the gap between the two decreases when we increase the complexity of the interaction terms, and would perhaps have overtaken the logistic regression model, had it been increased it further. The very good performance of penalised logistic regression without interaction terms also indicates that the data generating model may be well approximated with just main effects in this case.

In the second part of the comparison, data was simulated from 'Setting 2'.
In this case, the size of the components of the model was varied, as well as for the additive tree model, between 2, 3 and 4. For the logic
regression model, the optimal model was chosen as previously described.
The learning rate for the additive copula regression method was fixed at $\frac{1}{3}$. The corresponding results are shown in Tables \ref{tab:complik_2} and \ref{tab:compauc_2}. The additive copula regression model now performs significantly better than
the others in terms of both the likelihood and the AUC. 

The model cannot exactly recreate the data generating model, although it is also based on a vine copula, but only approximate it, and thus should not have an unfair advantage over the other methods.

Again, the size of the components of the model mainly affected the
likelihood, whereas the AUC was almost unchanged, at least when increasing the component size from 3 to 4. The model was also fitted with learning rates $\frac{2}{3}$ and $1$, but the results were so similar to the ones for learning rate $\frac{1}{3}$, that they are not shown here.

\section{Examples on real data}\label{sec:examples}
To further illustrate the modelling approach, a couple of 
examples on real data are considered.

\subsection{Wisconsin breast cancer data}
First, a dataset \citep{street1993nuclear} of 569 observations,
where each observation consists of 32 (continuous) covariates derived from an
image of a small amount of tissue from a breast tumour extracted with a very
fine needle, so called \textit{fine needle aspirations} (FNAs), and a binary
label indicating whether the tumour is benign or malignant is considered. The additive copula regression model is fit
using model components with $L=3$ covariates in each, to a training set
consisting of $427$ of the observations, and $142$ of the observations are kept as a test set to make predictions on. In the paper by
\citet{street1993nuclear}, they fit a multi-surface method tree using three of
the covariates, which is a type of neural network with one hidden layer that is
fitted using linear programming \citep{mangasarian1993mathematical}. Therefore a comparison is made with a neural network with one hidden layer to the model, that uses the same three covariates. The performance of the models is judged in terms of the  log-likelihood, the AUC, and the accuracy. The accuracy is defined as one minus the zero-one loss, and is reported instead of the
zero-one loss because \citet{mangasarian1993mathematical} uses this measure.

\begin{table}
 \caption{Summary of results of fits to the Wisconsin breast cancer dataset.
 \label{table:wisconsin}}
 \footnotesize
 \begin{tabular}{c|c|c|c|c|c|c}
\hline
 & copulaboost & ridge & randomforest & xgboost & Logic regression & Neural net model \\
\hline
\multicolumn{7}{c}{likelihood}\\
\hline
Training & -22.827 & -47.632 & -54.616 & -2.388 & -11.185 & -45.802\\
Test & -4.533 & -11.399 & -11.572 & -8.511 & -15.374 & -7.519\\
\hline
\multicolumn{7}{c}{AUC}\\
\hline
Training & 0.9979 & 0.9941 & 0.9892 & 1 & 0.9992 & 0.989\\
Test & 0.9996 & 0.9991 & 0.9976 & 0.9972 & 0.9963 & 0.9996\\
\hline\\
\multicolumn{7}{c}{accuracy}\\
\hline
Training & 0.9859 & 0.9813 & 0.9508 & 1 & 0.993 & 0.9649\\
Test & 0.9859 & 0.9648 & 0.9789 & 0.9648 & 0.9437 & 0.9859\\
\hline
\end{tabular}

\end{table}

In addition, models using ridge regression, xgboost,
random forest, and logic regression are fitted for comparison. In the latter case, the data are first
'dichotomised' by replacing each covariate with the best decision stump
using the training data. For the additive copula regression model, the xgboost model, and the logic
regression model, the values of the tuning parameters, that is, the number
of trees or copula models in the two former cases, and the number of trees and
leaves in the latter, are set by choosing the best model within the training set
according to either the likelihood, AUC, or the accuracy. For the ridge
regression model the value of the penalty parameter is set by 10-fold cross validation of the log-likelihood. The results of these model fits are given in Table \ref{table:wisconsin}. One can see that the additive copula regression model gave the best out-of-sample results according to all of the criteria, although the neural net model achieves an equally good result in terms of the out of sample AUC. 

\begin{table}
  \caption{Top 5 variable combinations, in terms of the increase in likelihood,
  shown in the rightmost column\label{table:varimp}}
  \begin{tabular}{| c | c | c | c |}
    \hline
    radius worst & texture worst & concave points worst & 0.468\\
    radius se & texture se & texture worst &  0.175\\
    radius se & radius worst & area worst & 0.117\\
    concavity mean & compactness se & concave points worst & 0.11041\\
    concave points se & radius worst & texture worst & 0.11040\\
    \hline
  \end{tabular}
\end{table}

An interesting point of investigation is to examine which variables are included
in each of the model components, and which of these gave the largest increase to
the likelihood, and compare these to the variable combination that was used in
the paper by \citet{street1993nuclear}. The top five of the former are shown in
Table \ref{table:varimp}. The variable combination that was found in the paper
by was mean texture, worst area and worst smoothness, which does not appear as
any of the most important components of our model, and only one of them appears
in the top five components. However, if one looks at the correlations between
these, the spearman correlation between these and the variables in our top
component, the correlation between the pairs \textit{radius worst, area worst}
and \textit{texture worst, texture mean}, are $0.999$ and $0.91$, respectively.
In addition, the pairs \textit{concave points worst, area worst} and 
\textit{concave points worst, smoothness worst} have a spearman correlation of
$0.774,$ and $0.544,$ respectively.

\subsection{Boston housing data}

A well-known dataset that first appeared in a paper
by \citep{harrison1978hedonic} is also considered, which contains information collected by the U.S. Census Service about housing and housing prices in Boston, Massachusetts.
There are fourteen variables in the dataset, that describe different quantities
pertaining to housing in a given area. One of these is the median value of
owner-occupied homes. An indicator of whether the median value is
above the $76$-th percentile is used as the outcome variable, as there seems to be a sudden drop in the density of the median value around this value, and the others as covariates from which we want to make predictions about the outcome. There are $506$ observations in the data set, of which $380$ are used for estimation, and $126$ observations are left out as a test set. 

\begin{table}[h]
 \caption{Summary of results of fits to the Boston housing dataset.
 \label{table:boston}}
 \begin{tabular}{c|c|c|c|c|c}
\hline
 & copulaboost & ridge & randomforest & xgboost & Logic regression \\
\hline
\multicolumn{6}{c}{likelihood}\\
\hline
Training & -74.481 & -85.962 & -18.977 & -8.627 & -53.774\\
Test & -20.119 & -29.01 & -23.37 & -21.122 & -35.017\\
\hline
\multicolumn{6}{c}{AUC}\\
\hline
Training & 0.9659 & 0.9529 & 1 & 1 & 0.9591\\
Test & 0.9779 & 0.9605 & 0.9746 & 0.9797 & 0.9704\\
\hline
\multicolumn{6}{c}{accuracy}\\
\hline
Training & 0.9395 & 0.9368 & 1 & 0.9395 & 0.9526\\
Test & 0.9206 & 0.9286 & 0.9524 & 0.9206 & 0.8968\\
\hline
\end{tabular}

\end{table}

The same types of models are fit as for the Wisconsin breast cancer data,
with the exception of the neural net model. Further, components of $L=5$ 
covariates are used for the additive copula regression model. The results of these model fits, both in and out of sample 
are given in Table \ref{table:boston}. One can see from the results that the additive copula regression model 
achieves the best fit, in terms of the likelihood out of sample, and almost as good 
a fit as xgboost in terms of the AUC, and better than the other methods. For the 
zero-one loss (reported in the table in terms of the accuracy), the out of sample performance is somewhat worse 
than for the other methods, except ridge regression. The difference is however 
very small as our model misclassified $8$ of the observations compared to the 
random forest model, that misclassified $6$, and xgboost and logic regression, 
that misclassified $7$.

\section{Concluding remarks}\label{sec:conclusion}
The authors have proposed a type of additive models that consist of components made from pair-copula constructions. The design of the model components
makes the model well suited for situations where there are interaction
effects present in the data generating mechanism. Further, they do not rely
on a discretisation of continuous covariates, such as in decision trees,
and are therefore adequate for problems with many such covariates.
Moreover, a fitting algorithm of the gradient boosting type was designed, that involves a specialised model selection procedure. Several efforts to make the fitting procedure as efficient as possible were also done, by imposing some constraints on the model space and by
using approximations to speed up time-costly computations. These models
are useful by themselves, but in addition, the authors believe that some of the ideas used to improve the efficiency of the model fitting procedure could be useful for designing more efficient models selection and fitting algorithms for other types of copula-based regression models, such as the one of \cite{noh2013copula}. 

Empirical margins are often commonly used in copula models, as they are
flexible and do not require model selection. However, a comparison with
a simple parametric marginal model through simulations indicated that a
parametric approach works best when the aim is prediction, as in this
setting. Further, the characteristics of the method were explored in a
simulation study. The results indicate that larger model components
tend to improve the AUC, but not necessarily the likelihood. As the
computational effort increases with the size of the components, a
component size of 3-4 may in many situations represent a good
compromise.

In addition the model was compared to natural alternatives like logic
regression and boosting with decision trees. In particular, how different proportions of discrete covariates affected the
performance of the method relative to other methods was explored. In the simulations, the boosted copula-based regression method gave results that were either better than or comparable to the other methods, even when all covariates were discrete. The approach was also illustrated on a set of breast cancer data and on the Boston housing data. In both cases, the model was among the best performing ones.

This paper has focused on a binary response, but, as mentioned
earlier, the model formulation makes it straightforward to extend
it to other exponential family responses, by substituting the link function.
Moreover, the optimal number of model components was chosen using a validation set, but that may not be an adequate solution if data are not abundant. In such cases, one might instead use
information criteria, for instance an adapted version of the criteria
presented in \citet{lunde2020information}, which could lead the
algorithm to make better model selection choices.

\section{Acknowledgements}
This work is funded by The Research Council of Norway centre Big Insight, Project 237718.

\bibliography{article2}

\begin{appendix}

\section{Additional tables from the simulation study}

\begin{table}[htbp]
  \centering
  \resizebox{\textwidth}{!}{\begin{minipage}{\textwidth}\label{tab:lik_initial}
        \caption{Results of simulation in terms of the likelihood, with the standard deviation in parentheses.}
        \begin{tabular}{c|c|c|c|c|c|c}
\hline
Learning rate & 1 Covariate & 2 Covariates & 3 Covariates & 4 Covariates & 5 Covariates & 6 Covariates \\
\hline
\multicolumn{2}{c}{Setting one}\\
\hline
0.33 & -201.84 (14.98) & -199.92 (13.7) & -199.7 (13.54) & -200.14 (13.5) & -200.13 (13.63) & -200.66 (13.71)\\
0.67 & -201.48 (14.22) & -199.86 (13.74) & -199.79 (13.73) & -199.73 (13.67) & -199.97 (13.6) & -200.36 (13.63)\\
1 & -201.51 (14.14) & -199.67 (13.66) & -199.72 (13.73) & -199.53 (13.73) & -199.98 (13.7) & -200.31 (13.64)\\
\hline
\multicolumn{2}{c}{Setting two}\\
\hline
0.33 & -243.65 (12.33) & -240.36 (11.78) & -239.54 (11.67) & -239.05 (11.61) & -238.81 (11.33) & -238.3 (11.42)\\
0.67 & -242.12 (12.13) & -240.14 (11.89) & -239 (11.59) & -238.71 (11.33) & -238.36 (11.21) & -238 (11.32)\\
1 & -241.78 (11.83) & -240.06 (11.84) & -239.31 (11.8) & -238.76 (11.25) & -238.27 (11.44) & -238.04 (11.48)\\
\hline
\multicolumn{2}{c}{Setting three}\\
\hline
0.33 & -264.04 (11.38) & -261.39 (10.75) & -261.25 (10.9) & -261.47 (11.08) & -261.3 (10.93) & -261.19 (10.91)\\
0.67 & -264.07 (11.73) & -262.47 (11.47) & -262.56 (11.58) & -263.37 (12.05) & -263.28 (11.65) & -263.04 (11.7)\\
1 & -264.42 (11.81) & -263.93 (12.32) & -263.77 (12.35) & -265.11 (12.89) & -264.86 (12.43) & -264.42 (12.39)\\
\hline
\multicolumn{2}{c}{Setting four}\\
\hline
0.33 & -296.25 (10.37)  & -293.84 (10.57)  & -293.32 (10.71)  & -293.48 (11.09)  & -293.44 (10.95)  & -293.57 (11.1) \\
0.67 & -311.11 (7.68) & -309.67 (7.80) & -308.91 (7.88) & -308.26 (7.93) & -307.53 (7.86) & -307.17 (7.89) \\
1 & -294.73 (10.34)  & -294.25 (11.27)  & -294.64 (11.69)  & -295.36 (12.06)  & -295.4 (12.57)  & -296.02 (12.54) \\
\end{tabular}

      \end{minipage}}
\end{table}

\begin{table}[htbp]
  \centering
  \resizebox{\textwidth}{!}{\begin{minipage}{\textwidth}\label{tab:auc_initial}
        \caption{Results of simulation in terms of the AUC, with the standard deviation in parentheses.}
        \begin{tabular}{c|c|c|c|c|c|c}
\hline
Learning rate & 1 Covariate & 2 Covariates & 3 Covariates & 4 Covariates & 5 Covariates & 6 Covariates \\
\hline
\multicolumn{2}{c}{Setting one}\\
\hline
0.33 & 0.8838 (0.0176) & 0.886 (0.0167) & 0.8866 (0.0168) & 0.8866 (0.0167) & 0.8869 (0.017) & 0.8875 (0.0171)\\
0.67 & 0.8844 (0.0178) & 0.8864 (0.0164) & 0.8872 (0.0168) & 0.8867 (0.0167) & 0.887 (0.0164) & 0.8875 (0.0169)\\
1 & 0.8843 (0.0176) & 0.8865 (0.0162) & 0.8872 (0.0167) & 0.8867 (0.017) & 0.8876 (0.0163) & 0.8873 (0.0169)\\
\hline
\multicolumn{2}{c}{Setting two}\\
\hline
0.33 & 0.8246 (0.0201) & 0.8271 (0.0197) & 0.8287 (0.0193) & 0.8305 (0.0189) & 0.8309 (0.0191) & 0.8314 (0.0197)\\
0.67 & 0.8256 (0.0192) & 0.8283 (0.0198) & 0.8296 (0.0199) & 0.8305 (0.0195) & 0.8318 (0.019) & 0.8316 (0.0194)\\
1 & 0.8262 (0.0197) & 0.8288 (0.0196) & 0.8293 (0.0195) & 0.8304 (0.019) & 0.8312 (0.0197) & 0.8315 (0.0189)\\
\hline
\multicolumn{2}{c}{Setting three}\\
\hline
0.33 & 0.7734 (0.0246) & 0.7788 (0.0228) & 0.7785 (0.0235) & 0.7797 (0.0231) & 0.7794 (0.0233) & 0.7798 (0.0228)\\
0.67 & 0.7745 (0.0243) & 0.7788 (0.0234) & 0.7795 (0.0232) & 0.7798 (0.0236) & 0.7793 (0.0231) & 0.7803 (0.0231)\\
1    & 0.7746 (0.0239) & 0.7788 (0.0230) & 0.7796 (0.0233) & 0.7800 (0.0230) & 0.7792 (0.0229) & 0.7802 (0.0233)\\
\hline
\multicolumn{2}{c}{Setting four}\\
\hline
0.33 & 0.6841 (0.0275) & 0.6915 (0.0264) & 0.6940 (0.0263) & 0.6950 (0.0271) & 0.6953 (0.0265) & 0.6955 (0.0265)\\
0.67 & 0.6765 (0.0324) & 0.6840 (0.0305) & 0.6895 (0.0290) & 0.6883 (0.0283) & 0.6880 (0.0285) & 0.6871 (0.0265)\\
1    & 0.6876 (0.0266) & 0.6929 (0.0262) & 0.6937 (0.0264) & 0.6942 (0.0262) & 0.6945 (0.0280)  & 0.6950 (0.0265)\\
\end{tabular}

      \end{minipage}}
\end{table}

\begin{table}[htbp]
  \centering
  \resizebox{\textwidth}{!}{\begin{minipage}{\textwidth}\label{tab:iter3}
        \caption{Chosen number of components in setting three.}
        \begin{tabular}{c|c|c|c|c|c|c}
\hline
Learning rate & 1 Covariate & 2 Covariates & 3 Covariates & 4 Covariates & 5 Covariates & 6 Covariates \\
\hline
\multicolumn{2}{c}{Likelihood}\\
\hline
0.33 & 743.44 (179.41) & 367.26 (99.57) & 228.17 (58.39) & 178.87 (53.64) & 133.97 (39.02) & 105.64 (27.64)\\
0.67 & 439.74 (198.25) & 219.11 (102.37) & 136.33 (57.45) & 103.33 (47.03) & 77.47 (33.46) & 62.41 (24.49)\\
1 & 289.06 (150.99) & 155.92 (86.42) & 97.98 (47.76) & 76.73 (42.6) & 57.14 (28.77) & 45.66 (21.73)\\
\hline
\multicolumn{2}{c}{AUC}\\
\hline
0.33 & 386.99 (355.33) & 262.95 (176.69) & 174.2 (99.41) & 151.01 (89.3) & 112.78 (62.78) & 87.64 (46.11)\\
0.67 & 282.41 (294.81) & 192.28 (159.3) & 127.98 (88.55) & 103.04 (83.7) & 79.71 (58.7) & 61.34 (42.8)\\
1 & 241.3 (258.41) & 151.31 (143.08) & 92.11 (82.56) & 74.38 (71.5) & 59.56 (55.32) & 45.86 (39.99)\\
\end{tabular}

      \end{minipage}}
\end{table}

\end{appendix}
\end{document}